\documentclass[showpacs,amsmath,amssymb,onecolumn,prx,superscriptaddress,nofootinbib]{revtex4-1}

\usepackage[dvips]{graphicx} % for figures
\usepackage{amsfonts}
\usepackage{amssymb}
\usepackage{amscd}
\usepackage{amsmath}    % need for subequations
\usepackage{enumerate}
\usepackage{epsfig}
\usepackage{subfigure}
\usepackage{xcolor}
\usepackage{amsthm}
\usepackage{slashbox}

\newcommand{\bra}[1]{\mbox{$\left\langle #1 \right|$}}
\newcommand{\ket}[1]{\mbox{$\left| #1 \right\rangle$}}

\begin{document}

\title{More Randomness from a Prepare-and-Measure Scenario with Independent Devices }

\author{Yun-Guang Han}
\address{Key Laboratory of Quantum Information, University of Science and Technology of China, Hefei 230026, China}
\address{Synergetic Innovation Center of Quantum Information $\&$ Quantum Physics, University of Science and Technology of China, Hefei, Anhui 230026, China}
\author{Zhen-Qiang Yin}
\email{yinzheqi@mail.ustc.edu.cn}
\address{Key Laboratory of Quantum Information, University of Science and Technology of China, Hefei 230026, China}
\address{Synergetic Innovation Center of Quantum Information $\&$ Quantum Physics, University of Science and Technology of China, Hefei, Anhui 230026, China}
\author{Hong-Wei Li}
\address{Key Laboratory of Quantum Information, University of Science and Technology of China, Hefei 230026, China}
\address{Synergetic Innovation Center of Quantum Information $\&$ Quantum Physics, University of Science and Technology of China, Hefei, Anhui 230026, China}
\author{Wei Chen}
\address{Key Laboratory of Quantum Information, University of Science and Technology of China, Hefei 230026, China}
\address{Synergetic Innovation Center of Quantum Information $\&$ Quantum Physics, University of Science and Technology of China, Hefei, Anhui 230026, China}
\author{Shuang Wang}
\email{wshuang@ustc.edu.cn}
\address{Key Laboratory of Quantum Information, University of Science and Technology of China, Hefei 230026, China}
\address{Synergetic Innovation Center of Quantum Information $\&$ Quantum Physics, University of Science and Technology of China, Hefei, Anhui 230026, China}

\author{Guang-Can Guo}
\affiliation{Key Laboratory of Quantum Information, University of Science and Technology of China, Hefei 230026, China}
\address{Synergetic Innovation Center of Quantum Information $\&$ Quantum Physics, University of Science and Technology of China, Hefei, Anhui 230026, China}
\author{Zheng-Fu Han}
\affiliation{Key Laboratory of Quantum Information, University of Science and Technology of China, Hefei 230026, China}
\address{Synergetic Innovation Center of Quantum Information $\&$ Quantum Physics, University of Science and Technology of China, Hefei, Anhui 230026, China}

\begin{abstract}
How to generate genuine quantum randomness from untrusted devices is an important problem in quantum information processing. Inspired by the previous work on self-testing quantum random number generator[Phys. Rev. Lett. 114, 150501], we present a new method to generate quantum randomness from a prepare-and-measure scenario with independent devices. In existing protocols, the quantum randomness only depends on a witness value (e.g., CHSH value ), which is calculated with the observed probabilities. Differently, here all the observed probabilities are directly used to calculate the min-entropy in our method. Through numerical simulation, we find that the min-entropy of our proposed scheme is higher than the previous work, when a typical untrusted BB84 setup is used. Consequently, thanks to the proposed method, more genuine quantum random numbers may be obtained than before.
\end{abstract}


\maketitle

\section{Introduction}

True randomness is an essential resource in quantum information processing and has multiple applications in numerical simulation, statistics, lottery games and cryptography. Since it is impossible to generate true random numbers by computer algorithms, most true random number generators are based on unpredictable physical process. Recently a variety of quantum random number generation (QRNG)schemes based on the intrinsic randomness of quantum theory have been proposed\cite{QRNG1994,QRNG2000_1,QRNG2000_2,QRNG2007,QRNG2008_1,QRNG2008_2,QRNG2009,QRNG2010_1,QRNG2010_2,QRNG2011_1,QRNG2011_2,QRNG2014_1,QRNG2014_2,QRNG2014_3}. All of these schemes work essentially according to the same principle, exploiting the randomness of quantum measurements. However, the random numbers generated by these protocols relies on the assumption of the specific internal functioning of devices. The output data can only be tested by statistical method, such as statistical test suite from NIST\cite{NIST}. The statistical method cannot guarantee the true randomness of the output data. Furthermore, if the devices are spoiled or controlled by an adversary, the output data may be just pseudo-random numbers. To solve this problem, ``Device-Independent" (DI) QRNG was built\cite{DIQRNG}, which does not need knowledge of the internal functioning of the devices. The private randomness in DI protocols is certified by Bell inequality violation but not the details of the quantum devices. Unfortunately, such protocols are quite impractical under current technology, since they demand the total efficiency must be very high to avoid detection loophole attacks. Inspired by the DI approach to true randomness, Li et al proposed the semi-device-independent random number generation protocol\cite{Li11}. Semi-device-independent approach works in a prepare-and-measure scenario in which no assumption is made on the internal functioning of the preparation and measurement devices, except that the dimension of the quantum system accessed by the measurement device is bounded\cite{SDI}. However, this protocol still suffers from detection loophole attacks\cite{SDIAtt}.

Last year, Bowles et al proposed a new scheme based on a prepare-and-measure setup \cite{BQB14} and experimentally realized it\cite{Exp15}. This protocol (BQB14 for abbreviation) seems like SDI protocol, but requires the assumptions that the preparation and measurement devices are independent and the quantum system has bounded dimension. This protocol uses a dimension witness value to characterize the quantum randomness of the system. Since the witness value is given by an equality, this protocol can be used to generate randomness with high channel loss.
Here we present a novel QRNG protocol also in a prepare-and-measure scenario with independent devices. The assumption of our protocol is completely the same as the BQB14.  We make no assumption on the functioning of the devices except the dimension of preparation device is set to be 2 and its hidden variables are independent of any other devices.  The key difference between our protocol and the BQB14 is that: we use all the observed probabilities instead of a witness value as the index of the potential  quantum randomness. In BQB14 and even all SDI, DI protocols, one must use the observed probabilities to calculate a witness value, then use this witness value to calculate the quantum randomness of the output data. Unlike the existing protocols, we search all the possible quantum preparation and measurement process satisfied all the observed probabilities to find the minimum real randomness of the output data.  The merit of our method is that all the observed probabilities are directly used to calculate the randomness, thus our method may be optimal than the existed protocols. Simulation results show our protocol works with very low detection efficiency. With a typical prepare-and-measure setup (untrusted BB84 setup\cite{BB84}), we find that entropy of the proposed protocol is higher than BQB14 protocol.

\section{Protocol}% \label{sec-our-protocol}

Our protocol can be implemented with standard BB84 QKD systems as Fig 1.

\begin{figure}[hbt]
\centering
\includegraphics[width=7.5cm]{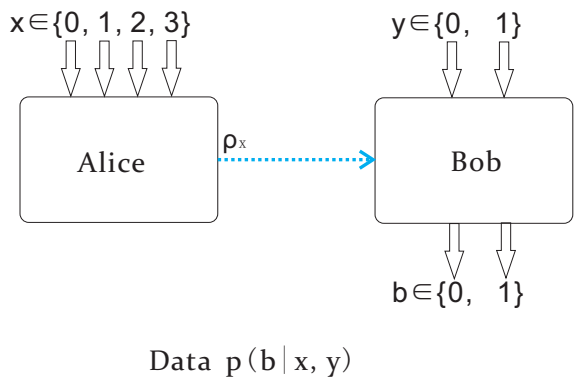}
\caption{Sketch of the protocol}
\label{Fig:protocol}
\end{figure}

The protocol is as follow:
\begin{enumerate}
  \item
    Alice randomly prepares four qubit states $\rho_{0}, \rho_{1}, \rho_{2}$, and $\rho_{3}$ by inputting x=0,1,2,3 to her device respectively.
  \item
  	For Bob there are two measurements y=0,1 with two outputs $b\in \{0,1\}$. In general, the measurement $M_y=\{M_y^0,M_y^1\}$ should be a POVM. However, we first assume that $M_y$ is a projective measurement for simplicity. The POVM case will be analyzed later in this paper.
  \item
    Alice and Bob observe the conditional probabilities $q(b|x,y)$. The task is to extract real quantum randomness generated by potential quantum process according to all the observed probabilities $q(b|x,y)$.
\end{enumerate}

Before proceeding, we must model our device with hidden variables. To model the characteristics of the preparation devices, we represent the internal state of the preparation device by a random variable $\lambda$. In each run of the experiment, the preparation device emits a qubit sate $\rho_x^\lambda$  which depends on the setting x and the internal hidden variable $\lambda$.  Hence when Alice inputs x, the device prepares $\sum_{\lambda}q_x^\lambda\rho_x^\lambda$. We assume that $\lambda$ is unknown to the legitimate users, and even also any adversary. This is the key difference between our model and DI protocols. In DI ones, the hidden variable $\lambda$ is planted by an adversary and thus known to the adversary. Conversely, in our model the adversary only knows the distribution of $\lambda$ but does not know the exact value of $\lambda$ of each run. Our assumption of preparation device is quite similar to BQB14. For measurement device, since we have assumed that the adversary including the measurement device has no idea about the exact value of $\lambda$ of each run , the measurement device performs an unknown measurement $M_y$, which is irrelevant of $\lambda$. As the observer has no access to the variable $\lambda$, he will only observe the distribution:
\begin{equation}
\begin{aligned}
q(b|x,y)&=\sum_{\lambda}q_x^\lambda p(b|x,y,\lambda)\\&=\sum_{\lambda}q_x^\lambda Tr(\rho_{x}^{\lambda}M_{y}^{b}),
\end{aligned}
\end{equation}

where
\begin{equation}
\rho_{x}=\sum_{\lambda}q_x^\lambda\rho_{x}^{\lambda}
\end{equation}

Without loss of generality, we can rewrite $\rho_x=\sum_{\lambda}q_x^\lambda\ket{\Psi^{\lambda}}\bra{\Psi^{\lambda}}$. The task of the legitimate user is to estimate the amount of genuine quantum randomness generated in the setup based only on the observed distribution $q(b|x,y)$. Since we have assumed that $M_y=\{M_y^0,M_y^1\}$ is a projective measurement, the genuine quantum randomness for the output data under measurement $M_y$ is given by maximum value of guess probability, which is $p_g=\sum_{\lambda}q_x^\lambda max\{\bra{\Psi^{\lambda}}M_y^0\ket{\Psi^{\lambda}},\bra{\Psi^{\lambda}}M_y^1\ket{\Psi^{\lambda}}\}$.
The maximum value of the guessing probability $p_g$ reflects the genuine quantum randomness. Since the hidden variable $\lambda$ is unknown to user, one should calculate $max_{\lambda}p_g$ by searching all possible distribution of $\lambda$ and decomposition of $\rho_x$.
A general consideration for how to calculate this value is given in the next section.

\section{Analysis}
We still consider the measurements are all projective measurements at first. For the output data with $\rho_x$ under measurement $M_y$, we define the maximum value of guessing probability as $max \ p_g(x,y)=\max\limits_{q_x^\lambda} \sum_{\lambda}q_x^\lambda \max\limits_{b\in\{0,1\}} Tr(\ket{\Psi_{x}^{\lambda}}\bra{\Psi_{x}^{\lambda}}M_y^b)$, which reflects the quantum randomness. Although the hidden variable $\lambda$ may have infinite values, we can be divided $\lambda$ into two parts by the value of $Tr(\ket{\Psi_{x}^{\lambda}}\bra{\Psi_{x}^{\lambda}}M_y^0)$ is higher than $\frac{1}{2}$ or not. So it will be not restrictive for the calculation of $max p_g(x,y)$ if we assume $\lambda$ can be just chosen from two values $\lambda_1$ and $\lambda_2$. The maximal guessing probability becomes
\begin{equation}
\begin{aligned}
max\  p_g(x,y) &=q_x^{\lambda1} max\ \{Tr(\ket{\Psi_{x}^{\lambda1}}\bra{\Psi_{x}^{\lambda1}}M_y^0),Tr(\ket{\Psi_{x}^{\lambda1}}\bra{\Psi_{x}^{\lambda1}}M_y^1)\}
\\&+q_x^{\lambda2} max\ \{Tr(\ket{\Psi_{x}^{\lambda2}}\bra{\Psi_{x}^{\lambda2}}M_y^0),Tr(\ket{\Psi_{x}^{\lambda2}}\bra{\Psi_{x}^{\lambda2}}M_y^1)\}
\end{aligned}
\end{equation}

The maximal guessing probability denotes the solution to the following optimization problem:
\begin{equation}
\begin{aligned}
&max\  p_g(x,y)\\
&subject\  to: \\
&\{q(b|x,y)=q_x^{\lambda1} Tr(\ket{\Psi_{x}^{\lambda1}}\bra{\Psi_{x}^{\lambda1}}M_y^b)+q_x^{\lambda2} Tr(\ket{\Psi_{x}^{\lambda2}}\bra{\Psi_{x}^{\lambda2}}M_y^b),b\in\{0,1\},x\in\{0,1,2,3\},y\in\{0,1\}\}
\end{aligned}
\end{equation}
Above formulae are based on the assumption that $M_y$ is a projective measurement. However, $M_y$ may be a POVM but not projective measurement.
Fortunately, as proved in \cite{POVM}, a general POVM can be decomposed into 3 different operations, which is performed randomly. Concretely, the three operations are performing a projective measurement to decide the output, or just generating 0 or 1 without any measurement. Then we will use this measurement model to estimate the quantum randomness. We can assume the probabilities to choose these three operations to be $\{m_y,u_y^0,u_y^1\} (m_y+u_y^0+u_y^1)$ for y=0 and 1 separately. As a result, the min-entropy for the output of measurement $M_y$ is given as
\begin{equation}
\begin{aligned}
max\  p_g(x,y) &=m_yq_x^{\lambda1} max\ \{Tr(\ket{\Psi_{x}^{\lambda1}}\bra{\Psi_{x}^{\lambda1}}M_y^0),Tr(\ket{\Psi_{x}^{\lambda1}}\bra{\Psi_{x}^{\lambda1}}M_y^1)\}
\\&+m_yq_x^{\lambda2} max\ \{Tr(\ket{\Psi_{x}^{\lambda2}}\bra{\Psi_{x}^{\lambda2}}M_y^0),Tr(\ket{\Psi_{x}^{\lambda2}}\bra{\Psi_{x}^{\lambda2}}M_y^1)\}+(1-m_y)
\end{aligned}
\end{equation}

Consider we are interested with two-dimension system, it is convenient to rewrite our formulae with Bloch vectors. The observed probabilities are rewritten as $q(0|x,y)=m_y(\frac{1}{2}+\frac{\vec{S}_x\cdot\vec{T}_y}{2})+u_y^0$, where $\vec{S}_{x}$ is the Bloch vector of the input state $\rho_x$, $\vec{T}_y$ is the Bloch vector of the projective measurement $M_y$. The problem of finding the genuine quantum randomness becomes the calculation of
\begin{equation}
\begin{aligned}
&max\  p_g(x,y)=max\ \{m_yq_x^{\lambda1}(\frac{1}{2}+\frac{1}{2}|\vec{S}_x^{\lambda1}\cdot\vec{T}_y|)
+m_yq_x^{\lambda2}(\frac{1}{2}+\frac{1}{2}|\vec{S}_x^{\lambda2}\cdot\vec{T}_y|)
+(1-m_y)\}\\
&subject\  to\\
&q(0|x,y)=m_y(\frac{1}{2}+\frac{1}{2}|\vec{S}_x\cdot\vec{T}_y|)+u_y^0\\
&\vec{S}_x=q_x^{\lambda1}\vec{S}_x^{\lambda1}+q_x^{\lambda2}\vec{S}_x^{\lambda2}
\end{aligned}
\end{equation}
where $q_x^{\lambda1}+q_x^{\lambda2}=1$, $\vec{S}_x^{\lambda1}$ and $\vec{S}_x^{\lambda2}$ are Bloch vectors for qubit states. This is an optimization problem of variables $m_y$, $u_y^b$, $q_x^{\lambda1}$, $q_x^{\lambda2}$, $\vec{S}_x$, $\vec{S}_x^{\lambda1}$, $\vec{S}_x^{\lambda2}$, $\vec{T}_y$ subject to above constraints.

In an experiment, Alice and Bob observe the probabilities $q(b|x,y)$ and then we can use numerical method to compute $max\  p_g(x,y)$. In practical, we are particularly interested in extracting randomness from an untrusted BB84 setup. In next section, we simplify the general result to be fit for the experimental results based on untrusted BB84 setup.

\section{Protocol in BB84 setup}

Now we consider how to realize our protocol with an untrusted BB84 implementation. Ideally, the input states for x=0,1,2,3 are $\{\ket{H},\ket{V},\ket{+},\ket{-}\}$ respectively, and the measurements for y=0,1 are projective measurements $\{\ket{H}\bra{H},\ket{V}\bra{V}\}$ and $\{\ket{+}\bra{+},\ket{-}\bra{-}\}$. The probability distribution $p(0|x,y)$ is shown in Table 1.
\begin{table}
  \centering
  \caption{Parameters .}
\begin{tabular}{ccccccc}  \hline  \hline
    $y\backslash x$ & $0$ & $1$ & $2$  & $3$  \\ \hline
    $0$ & $1-e_0$ & $e_1$ & $p(0|2,0)$ & $p(0|3,0)$ \\ \hline
    $1$ & $p(0|0,1)$ & $p(0|1,1)$ & $e_2$ & $1-e_3$\\
    \hline  \hline
  \end{tabular}
 \label{table1}
\end{table}

where $e_0$, $e_1$, $e_2$ and $e_3$ are quantum bit error rates(QBER).

The measurement results can be written in our measurement framework as:
\begin{equation}
p(0|0,0)=m_0(\frac{1}{2}+\frac{1}{2}|\vec{S}_0\cdot\vec{T}_0|)+u_0^0=1-e_0
\end{equation}

\begin{equation}
p(0|1,0)=m_0(\frac{1}{2}+\frac{1}{2}|\vec{S}_1\cdot\vec{T}_0|)+u_0^0=e_1
\end{equation}

\begin{equation}
p(0|2,0)=m_0(\frac{1}{2}+\frac{1}{2}|\vec{S}_2\cdot\vec{T}_0|)+u_0^0
\end{equation}

\begin{equation}
p(0|3,0)=m_0(\frac{1}{2}+\frac{1}{2}|\vec{S}_3\cdot\vec{T}_0|)+u_0^0
\end{equation}

\begin{equation}
p(0|2,1)=m_1(\frac{1}{2}+\frac{1}{2}|\vec{S}_2\cdot\vec{T}_1|)+u_1^0=1-e_2
\end{equation}

\begin{equation}
p(0|3,1)=m_1(\frac{1}{2}+\frac{1}{2}|\vec{S}_3\cdot\vec{T}_1|)+u_1^0=e_3
\end{equation}

The quantum bit error rate $e_0$, $e_1$, $e_2$ and $e_3$ can be measured in the experiment and are always between 0 and 0.5. We can find out the maximal value of mean guessing probabilities

\begin{equation}
\bar{P}=\frac{1}{4}max\ \sum_{x=0}^{3}p_g(x,0)
\end{equation}
i.e., the average guessing probability of the outcome for input states x=0,1,2,3 in the measurement y=0.

In an experiment base on untrusted BB84 setup, we may observe that $p(0|2,0)$ and $p(0|3,0)$  are close to 1/2, which means that $p(0|2,0)$ and $p(0|3,0)$ are possibly related to the mismatched basis events. Conversely, we may suspect that $p(0|0,0)$ and $p(0|1,0)$ are related to matched basis events. It is reasonable to extract more randomness from mismatched basis events than matched basis events. Hence, without loss of generality, we let  $p_g(0,0)=p_g(1,0)=1$ all the time and try to obtain a tighter upper bound for $p_g(2,0)+p_g(3,0)$.

As proved in the last section, to calculate $p_g(2,0)+p_g(3,0)$ we should decompose the input state $\rho_2$ and $\rho_3$ into two parts and search all over the qubit strategies to get the maximal guessing probability. The constraints can be simplified by some mathematical techniques. Considering the worst situation, equation (7) and (8) become

\begin{equation}
m_0+u_0^0\geq 1-e_0
\end{equation}

\begin{equation}
u_0^0\leq e_1
\end{equation}

by (11)-(12), we get

\begin{equation}
m_1*\frac{(\vec{S}_2-\vec{S}_3)\cdot\vec{T}_1}{2}=1-e_2-e_3
\end{equation}

Since $0<m_1\leq 1$, $0\leq |\vec{T}_1|\leq 1$, we get

\begin{equation}
|\vec{S}_2-\vec{S}_3|\geq 2(1-e_2-e_3)
\end{equation}

So the maximal guessing probability denotes the solution to the following optimization problem:
\begin{equation}
\begin{aligned}
&\bar{P}=\frac{1}{4}max\ \sum_{x=0}^{3}p_g(x,0)\leq \frac{1}{2}+\frac{1}{4}max\ \{p_g(2,0)+p_g(3,0)\} \\
&where\  p_g(x,y)=m_yq_x^{\lambda1}(\frac{1}{2}+\frac{1}{2}|\vec{S}_x^{\lambda1}\cdot\vec{T}_y|)
+m_yq_x^{\lambda2}(\frac{1}{2}+\frac{1}{2}|\vec{S}_x^{\lambda2}\cdot\vec{T}_y|)
+(1-m_y)\\
&subject\ to \\
&m_0+u_0^0\geq 1-e_0 \\
&u_0^0\leq e_1 \\
&|\vec{S}_2-\vec{S}_3|\geq 2(1-e_2-e_3) \\
&p(0|2,0)=m_0(\frac{1}{2}+\frac{1}{2}|\vec{S}_2\cdot\vec{T}_0|)+u_0^0 \\
&p(0|3,0)=m_0(\frac{1}{2}+\frac{1}{2}|\vec{S}_3\cdot\vec{T}_0|)+u_0^0 \\
&0<m_0,u_0^0\leq 0 \\
&q_x^{\lambda1}+q_x^{\lambda2}=1 (x=2,3)\\
&\vec{S}_x=q_x^{\lambda1}\vec{S}_x^{\lambda1}+q_x^{\lambda2}\vec{S}_x^{\lambda2}(x=2,3)
\end{aligned}
\end{equation}

Thus for observed QBERs, $p(0|2,0)$ and $p(0|3,0)$, we can calculate the maximal guessing probability numerically.

\section{Simulation}

In Fig 2, we plot the value of maximal guessing probability as a function of QBERs compared with BQB14 protocol.

\begin{figure}[hbt]
\centering
\includegraphics[width=7.5cm]{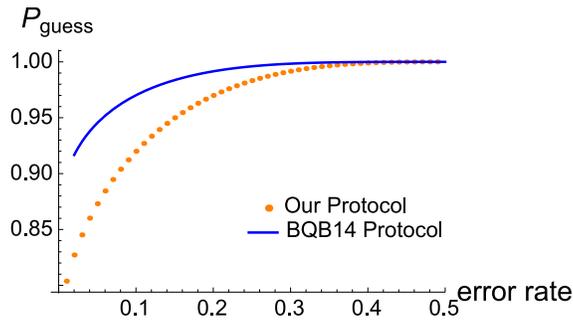}
\caption{Simulation: Maximal guessing probabilities vs QBERs. We set the four QBERs $e_0=e_1=e_2=e_3$ and $p(0|2,0)=p(0|3,0)=1/2$. The blue solid line is BQB14 protocol and the orange dashed line is our protocol.}
\label{Fig:simulation1}
\end{figure}

 In the simulation, we assume the four QBERs $e_0=e_1=e_2=e_3$ and $p(0|2,0)=p(0|3,0)=1/2$. From the simulation results, we can see both BQB14 protocol and our protocol can work in high noisy environment even when the QBERs are close to 0.5. And the maximal guessing probability in our protocol is lower than that in BQB14 protocol. In the ideal situation, the maximal guessing probabilities of our protocol is approximate to 0.75, and for BQB14 protocol, it is 0.854.

Then we use off-the-shelf experimental parameters to show the performance of the protocol in the presence of loss and noise, e.g., the loss is d dB, detection efficiency is $\eta_d=10\%$ and its dark count rate is $p_d=10^{-5}$. Besides, we consider a misalignment of detector $d_e=1\%$. Thus the overall QBER $e=\frac{0.5*(1-10^{-\frac{d}{10}})*p_d+\eta_d*d_e}{10^{-\frac{d}{10}}+(1-10^{-\frac{d}{10}})*p_d}$. And still $p(0|2,0)=p(0|3,0)=1/2$ .The simulation is shown in Fig 3. Results show our protocol can generate quantum randomness up to 25 dB.

\begin{figure}[hbt]
\centering
\includegraphics[width=7.5cm]{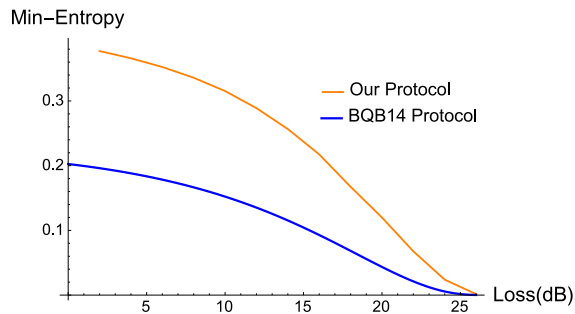}
\caption{Simulation: Min-entropy vs channel loss(dB). We set $\eta_d=0.1,d=10^{-5}$ per pulse. The detector error rate is $1\%$. The blue (lower) line is BQB14 protocol and the orange (upper) line is our protocol. }
\label{Fig:simulation2}
\end{figure}

\section{Discussion and Conclusion}

 Inspired by the pioneering work on quantum randomness generation \cite{BQB14,DIQRNG2014_1,DIQRNG2014_2} , we propose an alternative method which has higher quantumness generation rate at the same condition. Same as \cite{BQB14}, our method works in a prepare-and-measure scenario with independent devices. In our method, all observed probabilities are directly used to bound the min-entropy of the output data, while a specific witness value is used in other protocols. Hence, our method gives a tighter bound of min-entropy and thus, higher quantumness generation rate is obtained. Besides, our protocol maintains the advantage of BQB14 protocol that works in high lossy environment.

We use phase-randomized weak coherent source in experiment. However, our theory is for the single photon source. We provide two ways to overcome this problem. The first way is using a photon-number-resolving detector\cite{PNRD1, PNRD2}. Thus we can clearly distinguish single photon events from multiphoton events. Then we can discard all multi-photon events and just use the trials that correspond to single photon events to generate randomness. The second way is using the decoy states method when photon-number-resolving detector is not available. Similarly with decoy state quantum key distribution\cite{Decoy1,Decoy2,Decoy3}, we assume that Alice¡¯s source is phase-randomized weak coherent source. Then Alice can prepare additional decoy states besides the signal state by modulating the mean photon number of the laser pulses. In experiment, we observe that $q_\mu (b|x,y)$ directly, where $\mu$ is the mean photon number of the source. Note that $q_\mu(b|x,y)=\sum_{n=0}^{\infty} p_n(\mu) q_n(b|x,y)$, where $p_n(\mu)$ is the probability of n-photon events of a phase randomized weak coherent source, $q_n (b|x,y)$ is the probability of outputting b conditioned that the source emits a n-photon pulse, Alice inputs x and Bob inputs y. If we know $q_1 (b|x,y)$, we can calculate the min-entropy for single photon events with our theory. Then the min-entropy for all the events can be obtained by multiplying $p_1 (\mu)$, since we can assume the min-entropy for multi-photon events is 0. Fortunately, with the idea of decoy states we can establish some linear equations $q_\mu(b|x,y)=\sum_{n=0}^{\infty} p_n(\mu) q_n(b|x,y)$ by modulating different $\mu$.  Then the bounds of $q_1 (b|x,y)$ can be obtained by solving these linear equations. Furthermore, when the number of decoy states is infinite (modulating infinite different $\mu$), we can get the precise value of $q_1 (b|x,y)$ in principle. Then the calculation of min-entropy is straightforward by our theory. In conclusion, we can choose one from these two ways to exclude the effect of multiphoton events and generate true quantum randomness using our protocol.

\section*{Acknowledgments}
This work has been supported by the National Basic Research Program of China (Grants No. 2011CBA00200 and No. 2011CB921200), the National Natural Science Foundation of China (Grant Nos. 61475148, 61575183), and the ``Strategic Priority Research Program (B)" of the Chinese Academy of Sciences (Grant Nos. XDB01030100, XDB01030300).

\end{document}